\documentclass[twocolumn,prl,showpacs]{revtex4}
\usepackage{amsmath}
\usepackage{graphicx}

\newcommand{\eq}{\begin{equation}}
\newcommand{\fine}{\end{equation}}

\begin{document}

\title{Decoherence of a Macroscopic Quantum Superposition}
\author{Francesco De Martini$^{1,2}$, Fabio Sciarrino $^{3,1}$, and Nicol%
\`{o} Spagnolo$^{1}$}

\begin{abstract}
The high resilience to de-coherence shown by a recently discovered
Macroscopic Quantum Superposition (MQS)\ involving a number of photons in
excess of $5\times 10^{4}$ motivates the present theoretical and numerical
investigation. The results are placed in close comparison with the
properties of the well known MQS\ based on $\left| \alpha \right\rangle $
states. The very critical decoherence properties of the latter MQS are found
to be fully accounted for, in a direct a simple way, by a unique
''universal'' function: indeed a new property of the quantum ''coherent
states''.
\end{abstract}

\address{$^{1}$Dipartimento di Fisica dell'Universit\'{a} ''La Sapienza'' and Consorzio Nazionale Interuniversitario per le Scienze Fisiche della Materia, Roma, 00185 Italy\\
$^{2}$ Accademia Nazionale dei Lincei, via della Lungara 10, I-00165 Roma, Italy \\
$^{3}$Centro di Studi e Ricerche ''Enrico Fermi'', Via Panisperna 89/A,Compendio del Viminale, Roma 00184, Italy}

\maketitle

Since the early decades of Quantum Mechanics, the counter-intuitive
properties associated with the coherent superposition state of macroscopic
objects was the object of an intense debate epitomized by the celebrated
\textquotedblright Schr\"{o}dinger Cat\textquotedblright\ paradox \cite%
{Eins35,Schr35}. In particular, the actual feasibility of such quantum
object has always been tied to the alleged infinitely short persistence of
its quantum coherence, i.e. of its overwhelmingly rapid \textquotedblright
decoherence\textquotedblright . In modern times the latter property,
establishing a rapid merging of the quantum rules of microscopic systems
into the classical dynamics, has been interpreted as a consequence of the
entanglement between the macroscopic quantum system with the environment 
\cite{Niel00,Zure}. By discarding the environmental variables in the final
calculations the initially pure quantum state generally appears to decay
irreversibly towards a probabilistic classical mixture \cite{Dur02}.
Recently, the general interest for decoherence has received a renewed
interest in the framework of quantum information theory where it plays a
fundamental detrimental role\ since it conflicts with the experimental
realization of the quantum computer or of any quantum device bearing any
useful, relevant complexity \cite{Gori07}. In this respect a large
experimental effort has been devoted recently to the implementation of \
Macroscopic (i.e. many-particle) Quantum Superpositions states (MQS),
adopting photons, atoms and electrons in superconducting devices. Particular
attention has been devoted to the realization of the MQS involving
\textquotedblright coherent states\textquotedblright\ of light, which
exhibits interesting and elegant Wigner function representations \cite%
{Schl01}. The most notable results of this experimental effort have been
reached with atoms interacting with microwave fields trapped inside a cavity 
\cite{Brun92,Raim01} or for freely propagating fields \cite{Ourj06}.\
However, in spite of the long lasting efforts spent in these endeavors, in
these realizations the MQS\ has always proved to be so fragile that even the
loss of a single particle\ was found to be able to spoil any possibility of
a direct observation of its quantum properties. Precisely on the basis on
these negative results in many scientific communities, and also within very
influential editorial teams, grew the opinion that the \textquotedblright
Schr\"{o}dinger Cat\textquotedblright\ is indeed a ill defined, abstruse and
then avoidable concept since\ it fundamentally lacks of any directly
observable property \cite{Dur02}.

\begin{figure}[t]
\centering
\includegraphics[width=0.38\textwidth]{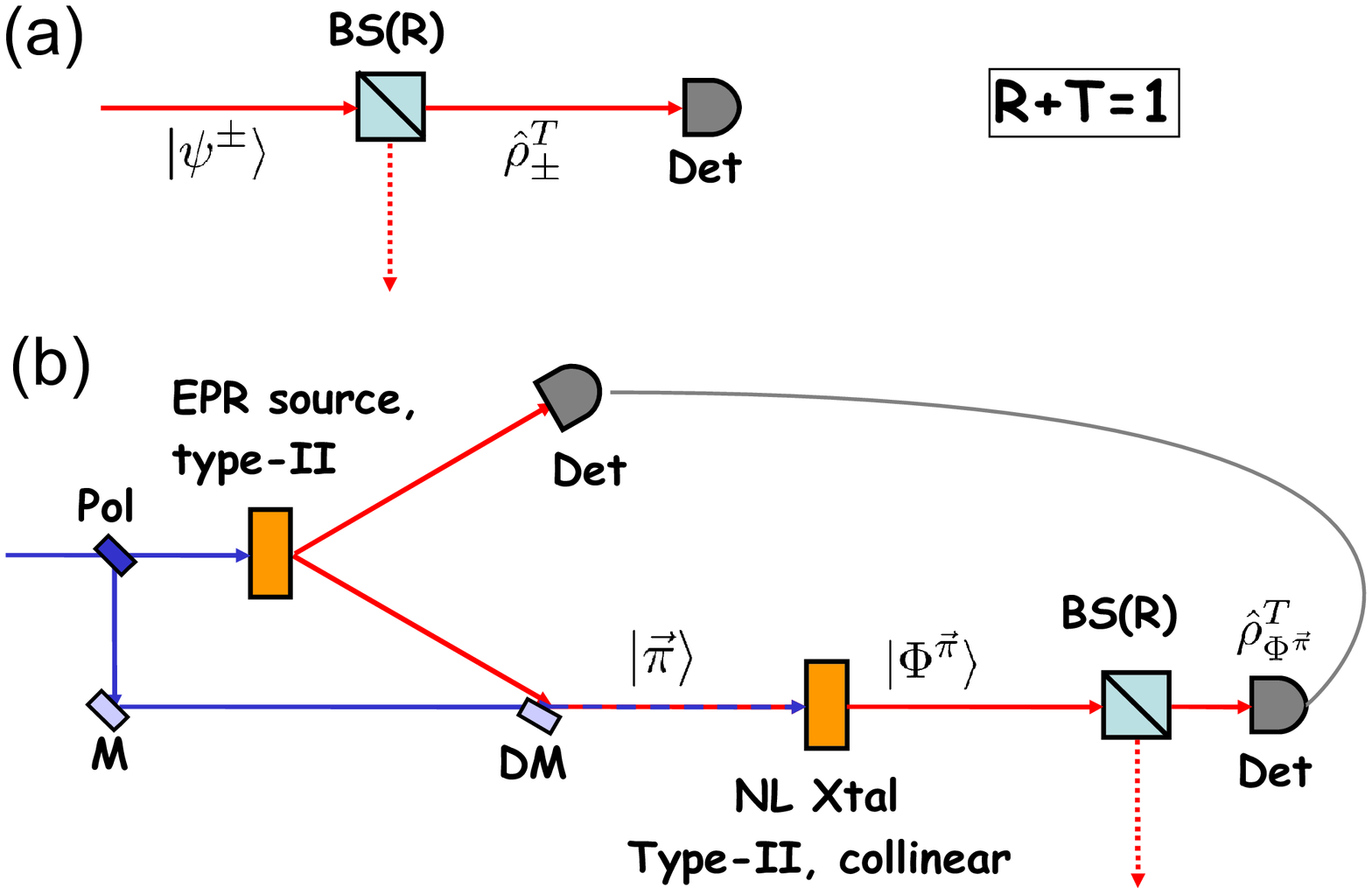}
\caption{Schematization of the decoherence model by a linear beam-splitter
of transmittivity T. (a) Analysis of quantum superposition of coherent
states $|\protect\psi ^{\pm }\rangle =\frac{1}{\protect\sqrt{N_{\pm }}}%
\left( |\protect\alpha \rangle \pm |-\protect\alpha \rangle \right) $. (b)
Analysis of amplified states of the collinear optical parametric amplifier
injected by a single photon qubit generated in a type-II EPR source.}
\label{fig:fig1}
\end{figure}

However, very recently a new kind of MQS\ involving a number of particles $N$
in excess of $5\times 10^{4}$ has been realized that allows the direct
observation of entanglement between a microscopic and a macroscopic photonic
state and shows a very high resilience to decoherence by coupling with
environment \cite{DeMa08}. Precisely, the MQS was generated by a
quantum-injected optical parametric amplifier (QI-OPA) seeded by a
single-photon belonging to an EPR entangled pair:\ indeed a high-gain
phase-covariant Cloning Machine \cite{DeMa98,DeMa05,DeMa05b,Scia05,Naga07}.\ By
this device, that includes an Orthogonality Filter (O-Filter, OF) for better
state discrimination, the microscopic-macroscopic state non-separability was
successfully tested and the microscopic-macroscopic Violation of the
Bell'inequalities for Spin-1 excitations was attained \cite{DeMa08,DeMa08b}.
It then appears that several prejudices about MQS should need a revision
and\ that a careful analysis looks necessary in order to learn more about
the MQS dynamics and its decoherence. The present work is precisely intended
to provide a first glimpse in this elusive albeit fundamental matter.

\textbf{Criteria for macroscopic superposition: } In order to distinguish
between two different quantum states, let us introduce a definition of
distance in the Hilbert space. An useful parameter to characterize
quantitatively the overlap of two quantum states is the\ ''fidelity'' $%
\mathcal{F\ }$between two generic density matrices $\widehat{\rho }$ and $%
\widehat{\sigma }$, defined as: $\mathcal{F}(\widehat{\rho },\widehat{\sigma 
})=\mathrm{Tr}\left( \sqrt{\widehat{\rho }^{\frac{1}{2}}\widehat{\sigma }%
\widehat{\rho }^{\frac{1}{2}}}\right) \ $and $0\leq \mathcal{F}\leq 1$,
where $\mathcal{F}=1$ for identical states, and $\mathcal{F}=0$ for
orthogonal states\cite{Jozs94}. This quantity is still not a metric in the
quantum state space, but can be adopted to define different useful metrics.
Amongst them we consider here the ''Bures distance'' \cite{Bure69,Hubn92}: 
\begin{equation}
D(\widehat{\rho },\widehat{\sigma })=\sqrt{1-\mathcal{F}(\widehat{\rho },%
\widehat{\sigma })}
\end{equation}
In this paper we shall adopt this ''distance'' as it is connected to the
probability of obtaining an inconclusive result with a suitable Positive
Operator Valued Measurement (POVM) \cite{Gisi96}, which is $\mathcal{F}%
\left( |\phi \rangle ,|\psi \rangle \right) =|\langle \psi |\phi \rangle |$
for pure states. When dealing with the distance amongst two MQS, we shall
also refer to $D$ as the MQS\ ''Visibility''.

\textbf{Distinguishability, MQS\ Visibility, the lossy Channel}\textit{.}%
\newline
Let us characterize two macroscopic states $|\phi _{1}\rangle $ and $|\phi
_{2}\rangle $ and the corresponding MQS's: $|\phi ^{\pm }\rangle =\frac{%
\mathcal{N}_{\pm }}{\sqrt{2}}\left( |\phi _{1}\rangle \pm |\phi _{2}\rangle
\right) $ by adopting two criteria. \textbf{I)} The \textit{%
distinguishability} between $|\phi _{1}\rangle $ and $|\phi _{2}\rangle $
can be quantified as $D\left( |\phi _{1}\rangle ,|\phi _{2}\rangle \right) $%
. \textbf{II)} The ''\textit{Visibility''}, i.e. ''degree of orthogonality''
of the MQS's $|\phi ^{\pm }\rangle $ be expressed again by: $D\left( |\phi
^{+}\rangle ,|\phi ^{-}\rangle \right) $. Indeed, the value of the MQS\
visibility depends exclusively on the relative phase of the component states:%
$\ |\phi _{1}\rangle $ and $|\phi _{2}\rangle $. Assume two orthogonal
superpositions $|\phi ^{\pm }\rangle $: $D\left( |\phi ^{+}\rangle ,|\phi
^{-}\rangle \right) =1$. In presence of losses the relative phase between $\
|\phi _{1}\rangle $ and $|\phi _{2}\rangle $ progressively randomizes and
the superpositions $|\phi ^{+}\rangle $ and $|\phi ^{-}\rangle $ approach an
identical fully mixed state leading to: $D\left( |\phi ^{+}\rangle ,|\phi
^{-}\rangle \right) =0$. The aim of this paper is to study the evolution in
a lossy channel of the phase decoherence acting on two macroscopic states $%
|\phi _{1}\rangle $ and $|\phi _{2}\rangle \ $and on the corresponding
superpositions $|\phi ^{\pm }\rangle $ and the effect on the size of the
corresponding $D\left( |\phi _{1}\rangle ,|\phi _{2}\rangle \right) $ and $%
D\left( |\phi ^{+}\rangle ,|\phi ^{-}\rangle \right) $.

\begin{figure}[t]
\caption{(a)-(d): Plot of the distribution of the number of photons in the $|%
\protect\psi ^{+}\rangle $ state for $\protect\alpha =4$, corresponding to
an average number of photons $\langle n\rangle =16$, for reflectivities $R=0$
(fig.2-a), $R=0.1$ (fig.2-b), $R=0.5$ (fig.2-c) and $R=0.8$ (fig.2-d). (e):
Plot of the universal curve that describes the distance between $|\protect%
\psi ^{+}\rangle $ and $|\protect\psi ^{-}\rangle $ after losses as a
function of $R\langle n\rangle \sin^{2} \protect\varphi$. }
\label{fig:fig2}\includegraphics[scale=.24]{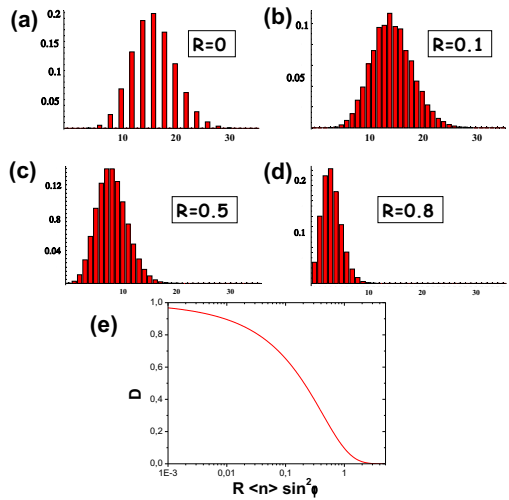}
\end{figure}

The effects of losses are analyzed through the effect of a generic linear
beam-splitter (BS)\ with transmittivity $T$ and reflectivity $R=1-T$ acting
on a generic quantum state associated with a single mode beam: Fig.\ref%
{fig:fig1} \cite{Loud}. The procedure for the calculation of the output
density matrix is the insertion of the BS\ unitary transformations as a
function of the parameters $T$ or $R$, and the evaluation of the partial
trace of the emerging field on the reflected mode (R-trace), i.e. on the
loss variables. In this paper we shall study quantitatively how the
distinguishability of macroscopic quantum states and the corresponding MQS\
Visibility is affected by the parameters $R$ or $T$ of the lossy channel. We
are going to carry out this study for two classes of states that bear a
particular relevance in the present context, as discussed above in the
introductory comments.\newline

\textbf{Quantum superposition of coherent states.} The first class of states
we are going to analyze is the quantum superposition of coherent states $%
|\phi _{1,2}\rangle =|\pm \alpha \rangle $, defined as: $|\psi ^{\pm
}\rangle =\frac{\mathcal{N}_{\pm }}{\sqrt{2}}\left( |\alpha \rangle \pm
|-\alpha \rangle \right) $\cite{Schl91}. As it is well known, this MQS
exhibits several very interesting quantum properties, such as squeezing and
sub-poissonian statistics and has been so far considered as the paradigmatic
representation of the Schr\H{o}dinger's cat State.

Since the BS\ doesn't affect the poissonian character on the field, the
application of the losses model to the input component coherent states $|\pm
\alpha \rangle $ of opposite phase leads again to an output coherent-state
density matrix of the form $\widehat{\rho }_{\alpha }^{T}=|\sqrt{T}\alpha
\rangle \,\langle \sqrt{T}\alpha |$. Then the distance between the two
states with opposite phase is: $D\left( |\sqrt{T}\alpha \rangle ,|-\sqrt{T}%
\alpha \rangle \right) =\sqrt{1-e^{-2T|\alpha |^{2}}}$, a value close to 1
except for $T$ $=0$, i.e. for total loss of all particles. Hence the
coherent states $|\pm \alpha \rangle $ keep their mutual distinguishability
through the lossy channel.

Let's now consider the MQS Visibility. Applying the previous losses
procedure to the $|\psi ^{\pm }\rangle $ states, the density matrices after
losses have the general form:

{\small 
\begin{equation}
\hat{\rho}_{C}^{\pm }=\frac{1}{2}\left( |\beta \rangle _{{}}\,_{{}}\langle
\beta |+|-\beta \rangle _{{}}\,_{{}}\langle -\beta |\pm e^{-2R|\alpha
|^{2}}\left( |-\beta \rangle _{{}}\,_{{}}\langle \beta |+|\beta \rangle
_{{}}\,_{{}}\langle -\beta |\right) \right) 
\end{equation}%
}

with $|\beta \rangle =|\alpha \sqrt{T}\rangle $. For the coherent state MQS
with no losses ($T=1$), the distribution in the Fock space exhibits only
elements with an even number of photons for $|\psi ^{+}\rangle $ or an odd
number of photons for $|\psi ^{-}\rangle $. This very peculiar \textit{comb}
structure is indeed very fragile under the effect of losses since the\
R-trace operation must be carried out in the space of the \textit{%
non-orthogonal }coherent-states. This is shown in Fig.\ref{fig:fig2}(a)-(b).
Furthermore, Fig.\ref{fig:fig2}(c)-(d) shows the effects of decoherence on
the $|\psi ^{\pm }\rangle $ states for increasing particle loss. The MQS\
Visibility of these states has been evaluated analytically in closed form
and is found extremely sensitive to decoherence \cite{nota}:

\begin{equation}
D=\sqrt{1-\sqrt{1-e^{-4R|\alpha |^{2}}}}
\end{equation}

Note that $D(x)$ depends only on the average number of \textit{lost }photons 
$x\equiv R|\alpha |^{2}$ $=$ $R<n>$, for \textit{any }value of $<n>$. The
average loss of 1 photon leads to the MQS\ Visibility value: $D=0.096$, and
then to the practical cancellation of any detectable interference effects
involving $\widehat{\rho }_{\psi ^{\pm }}^{T}$.\ This is fully consistent
with the experimental observations \cite{Brun92,Raim01}. The previous
calculations can be generalized to the general coherent state MQS: $|\psi
_{\varphi }^{\pm }\rangle {}$= $\frac{\mathcal{N}_{\varphi }^{\pm }}{\sqrt{2}%
}\left( |\alpha e^{\imath \varphi }\rangle \pm |\alpha e^{-\imath \varphi
}\rangle \right) $ leading respectively to the \textit{distinguishability}: $%
D(|\alpha e^{\imath \varphi }\rangle {},|\alpha e^{-\imath \varphi }\rangle
{})=\sqrt{1-e^{-2T|\alpha |^{2}\sin ^{2}\varphi }}$ and to the \textit{MQS
Visibility}:$\ D(|\psi _{\varphi }^{+}\rangle ,|\psi _{\varphi }^{-}\rangle )
$= $\sqrt{1-\sqrt{1-e^{-4R|\alpha |^{2}\sin ^{2}\varphi }}}$. We obtain the
previous results by replacing in $D(x)$ the coordinate $x$ by the rescaled
quantity: $x=R|\alpha |^{2}\sin ^{2}\varphi $. \ We can summarize the
theoretical results for the MQS\ Visibility by tracing the \textit{unique}
function:\ $D(|\psi _{\varphi }^{+}\rangle {},|\psi _{\varphi }^{-}\rangle
{})=D(x)$, shown in Fig. 2 (e)$.$We consider this \textquotedblright
universal\textquotedblright\ function an additional important property of
the \textquotedblright coherent states\textquotedblright , not previously
discovered to the best of our knowledge. Note that the function $D(x)$
approaches its minimum value\ with zero \textit{slope}: $Sl=\lim_{R%
\rightarrow 1}\left\vert dD(x)/dx\right\vert =0.$%
\begin{figure}[t]
\includegraphics[scale=.90]{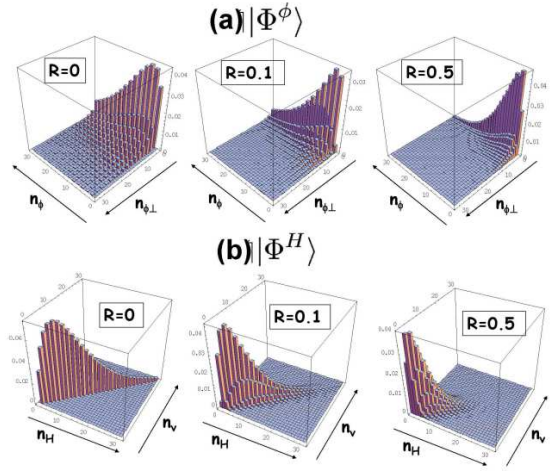}
\caption{(a) Probability distribution in the Fock space $(n_{\protect\phi %
},n_{\protect\phi _{\bot }})$ for the amplified $|\Phi ^{\protect\phi %
}\rangle $ state of a generic equatorial qubit for different values of the
transmittivity. (b) Probability distribution in the Fock space $(n_{H},n_{V})
$ for the amplified $|\Phi ^{H}\rangle $ state for different values of the
transmittivity. All distributions refer to a gain value of $g=1.5$,
corresponding to an average number of photons $\langle n\rangle \approx 19$. 
}
\label{fig:fig3}
\end{figure}

\textbf{Quantum superposition by phase-covariant quantum cloning.} The lossy
channel method has been applied to the amplified single photon qubits by a
collinear QI-OPA. The interaction Hamiltonian of this process is: $\widehat{%
\mathcal{H}}_{coll}=\imath \hbar \chi \widehat{a}_{H}^{\dag }\widehat{a}%
_{V}^{\dag }+\mathrm{h.c.}$ in the $\left\{ \vec{\pi}_{H},\vec{\pi}%
_{V}\right\} $ polarization basis, and $\widehat{\mathcal{H}}_{coll}=\frac{%
\imath \hbar \chi }{2}e^{-\imath \phi }\left( \widehat{a}_{\phi }^{\dag
\,2}-e^{\imath 2\phi }\widehat{a}_{\phi _{\bot }}^{\dag \,2}\right) $ for
any ''equatiorial'' basis $\left\{ \vec{\pi}_{\phi },\vec{\pi}_{\phi \perp
}\right\} $ on the Poincar\'{e} sphere where: $\vec{\pi}_{\phi }=\frac{1}{%
\sqrt{2}}\left( \vec{\pi}_{H}+e^{\imath \phi }\vec{\pi}_{V}\right) $. Two
relevant equatorial basis are $\left\{ \vec{\pi}_{+},\vec{\pi}_{-}\right\} $
and $\left\{ \vec{\pi}_{R},\vec{\pi}_{L}\right\} $ corresponding
respectively to $\phi =0$ and $\phi =\pi /2$. We remind that the \textit{%
phase-covariant} cloning process amplifies identically all ''equatorial''
qubits. The symbols H and V refer to horizontal and vertical field
polarizations, i.e. the extreme ''poles'' of the Poincar\'{e} sphere. By
direct calculation, the amplified states for an injected qubit $\pi =\left\{
H,V\right\} $ is: 
\begin{equation}
|\Phi ^{\pi }\rangle =\frac{1}{C^{2}}\sum_{i=0}^{\infty }\Gamma ^{i}\sqrt{i+1%
}\,|(i+1)\pi ,i\pi _{\bot }\rangle
\end{equation}
While for an injected equatorial qubit the amplified state is: 
\begin{equation}
|\Phi ^{\phi }\rangle =\sum_{i,j=0}^{\infty }\gamma _{ij}|(2i+1)\phi
,(2j)\phi _{\bot }\rangle  \label{eq:Phi_equat}
\end{equation}
where $\gamma _{ij}=\frac{1}{C^{2}}\left( \frac{e^{-\imath \varphi }\Gamma }{%
2}\right) ^{i}\left( -e^{\imath \varphi }\frac{\Gamma }{2}\right) ^{j}\frac{%
\sqrt{(2i+1)!}\,\sqrt{(2j)!}}{i!j!}$. In these expressions $C=\cosh g$ and $%
\Gamma =\tanh g$, where g is the non linear gain of the amplifier.

Let us consider the MQS of the macrostates $|\Phi ^{+}\rangle $ and $|\Phi
^{-}\rangle $: $|\Psi ^{\pm }\rangle =\frac{\mathcal{N}_{\pm }}{\sqrt{2}}%
\left( |\Phi ^{+}\rangle \pm i|\Phi ^{-}\rangle \right) $. Due to the
linearity of the amplification process \cite{DeMa05}, it can be easily found
that $|\Psi ^{\pm }\rangle =\left| \Phi ^{R/L}\right\rangle .$ The
distribution in the Fock space $P(n_{\Phi ,}n_{\Phi \perp })$ corresponding
to each ''equatorial'' macrostate $|\Phi ^{\phi }\rangle $, evaluated by Eq.(%
\ref{eq:Phi_equat}) exhibits a comb structure similar to the one shown in
Fig.2 (a). Indeed only terms with a specific parity, in particular with odd
number of photons for $\vec{\pi}_{\phi }$ polarization and even number of
photons for its orthogonal $\vec{\pi}_{\phi _{\bot }}$ are non vanishing.
Furthermore, the amplified $\left\{ \left| \Phi ^{H,V}\right\rangle \right\} 
$ states are characterized by a diagonal distribution. 

\begin{figure}[t]
\centering
\includegraphics[scale=.38]{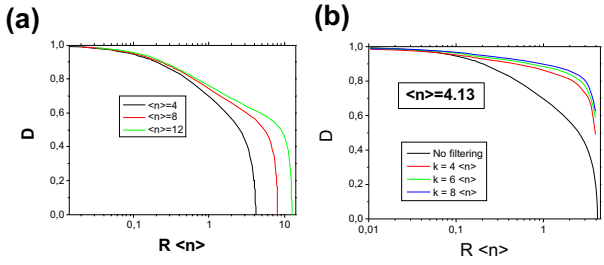}
\caption{(\textbf{a}) Numerical evaluation of the distance $D(x)$ between
two orthogonal equatorial macro-qubits $|\Phi^{\protect\phi,\protect\phi%
_{\bot}}\rangle$ as function of the average lost particle $x=R<n>$. Black
line refers to $g=0.8$ and hence to $\langle n \rangle \approx 4$, red
line to $g=1.1$ and $\langle n \rangle \approx 8$, green line to $g=1.3$
and $\langle n \rangle \approx 12$. \textbf{b}) Numerical evaluation of the Bures
distance between two orthogonal equatorial macro-qubits for different values
of the threshold $\overline{\protect\xi}$ ($g=0.8$). }
\label{fig:fig4}
\end{figure}

In fig.\ref{fig:fig3} the effects of losses in the distribution of both
equatiorial and $\left\{ \left| \Phi ^{H,V}\right\rangle \right\} $
amplified states are shown for different values of $T$. The peculiar
features of these Fock space distributions are progressively smoothed by the
effect of losses.

As a following step, we have evaluated numerically the \textit{%
distinguishability of }$\left\{ |\Phi ^{+,-}\rangle \right\} $ through the
distance $D(|\Phi ^{+}\rangle ,|\Phi ^{-}\rangle )$ between the states:
Fig.4-(\textbf{a}). It is found that this property of $\left\{ |\Phi
^{+,-}\rangle \right\} $ coincides with the MQS\ Visibility of $|\Psi ^{\pm
}\rangle $, in virtue of the phase-covariance of the process: $D(|\Psi
^{+}\rangle ,|\Psi ^{-}\rangle )$= $D(|\Phi ^{R}\rangle ,|\Phi ^{L}\rangle )$%
= $D(|\Phi ^{+}\rangle ,|\Phi ^{-}\rangle )$. The \textit{visibility} of the
MQS $\left\{ |\Psi ^{+,-}\rangle \right\} $ has been evaluated numerically
analyzing $D(x)$ as a function of the average lost photons: $x\equiv R<n>$.
The results for different values of the gain are reported in Fig.\ref%
{fig:fig4}-(\textbf{a}). Note that for small values of $x$ the decay of $D(x)
$ is far slower than for the coherent state case shown in Fig. 2-(\textbf{e}%
). Furthermore, after a common inflexion point at $D\sim 0.5$ the\ function $%
D(x)$ drops to zero for $R=1$ and then for:  $<n>$ $\sim N$, the total
number of particles in the primary beam. Very important, for large $<n>$,
i.e. $R\rightarrow 1$ the  \textit{slope} of the functions $D(x)$ increase
fast towards a very large value: $R\rightarrow 1$: $Sl=\lim_{R\rightarrow
1}\left\vert dD(x)/dx\right\vert \approx \infty $. All this means that the
MQS\ Visibility can be large even if the average number $x$ of lost
particles is close to the total number $N$, i.e. for $R\sim 1$. As seen,
this behavior is opposite to the case of \textit{coherent states} where the
function $D(x)$ approaches zero value with \textit{zero} slope: Fig.2(e). We
believe that this lucky and quite unexpected behavior is at the core of the
high resilience to decoherence of our QI-OPA MQS\ solution. Note that this
behavior was responsible for the well resolved interference pattern with
visiblity: $V\approx 20\%$ obtained in absence of  O-Filter (OF) by:\cite%
{Naga07}\  

\textbf{Orthogonality Filter} The demonstration of microscopic-macroscopic
entanglement by adopting the O-Filter was reported in \cite{DeMa08}. The
POVM like technique \cite{Pere95} implied by this device locally selects the
events for which the difference between the photon numbers associated with
two orthogonal polarizations $|m-n|>k$, i.e. larger than an adjustable
threshold, $k$ \cite{Naga07}. By this method a sharper discrimination
between the output states $|\Phi ^{\varphi }\rangle $ e $|\Phi ^{\varphi
_{\bot }}\rangle $ can be achieved. The action of the OF can be formalized
through the measurement operator $\hat{P}_{OS}=\sum_{m,n}|m+,n-\rangle
\,\langle m+,n-|$, where the sum over $m,n$ extends over the terms for which
the above inequality holds. In Fig.\ref{fig:fig4} the results of a numerical
analysis carried out for $g=0.8$ and different values of $k$\ are reported.
Note the the increase of the value of $D(x)$, i.e. of the MQS\ Visibility,
by increasing $k$ and, again: $Sl=\lim_{R\rightarrow 1}\left\vert
dD(x)/dx\right\vert \approx \infty $. Of course here the high
interference visibility is achieved at the cost of a lower success
probability, as expected.

The present work was intended to give a firm theoretical basis to the very
high resilience to decoherence demonstrated in recent experiments by our
QI-OPA\ generated Macroscopic Quantum Superposition. This novel MQS\ system
is expected to play a relevant role in the future investigations on the
Foundational structure of Quantum Mechanics. We acknowledge useful
discussions with Chiara Vitelli. Work supported by PRIN 2005 of MIUR and
INNESCO 2006 of CNISM.


\begin{references}
\bibitem{nota} The closed-form calculations leading to Eq.3 as well as similar expression for $D(x)$ will be reported in a forthcoming paper.
\end{references}
\end{document}